\documentstyle[preprint,prd,eqsecnum,aps,psfig]{revtex}\tighten

\newcommand{\beq}{\begin{equation}}
\newcommand{\beqa}{\begin{eqnarray}}
\newcommand{\eeq}{\end{equation}}
\newcommand{\eeqa}{\end{eqnarray}}
\newcommand{\p}{\phi}
\newcommand{\k}{\kappa}

\newcommand{\simg}{\gtrsim}
\newcommand{\siml}{\lesssim}

\begin{document}
\draft
\preprint{UTAP-321, gr-qc/9903094}

\title{
Quintessence, the Gravitational Constant, and Gravity
}
\author{
Takeshi Chiba
}
\address{
Department of Physics, University of Tokyo,
Tokyo 113-0033, Japan
}

\date{\today}

\maketitle

\begin{abstract}

Dynamical vacuum energy or quintessence, a slowly varying and 
spatially inhomogeneous component of the energy density with negative 
pressure, is currently consistent with the observational data. 
One potential difficulty with the idea of quintessence is that
couplings to ordinary matter should be strongly suppressed so as not
to lead to observable time variations of the constants of nature. 
We further explore the possibility of an explicit coupling between 
the quintessence field and the curvature. Since such a scalar field
gives rise to another gravity force of long range ($\simg H^{-1}_0$), 
the solar system experiments put a constraint
on the non-minimal coupling: $|\xi| \siml  10^{-2}$.  

\end{abstract}


\section{introduction}

Recently a number of observations suggest that 
 the Universe is dominated by an energy component with an
effective negative pressure\cite{os}. One possibility for such a component
is the cosmological constant. Another possibility is dynamical
vacuum energy or quintessence, a slowly varying and spatially
inhomogeneous component with negative 
pressure\cite{rp,fhw,xmatter,tw,cds,bs}.

We face two problems when we consider such a nonzero vacuum energy. 
The first is the fine-tuning problem related to the energy scale of 
the vacuum energy density $\sim 10^{-47}{\rm GeV}$. The second is the
coincident problem: why the vacuum energy is beginning to dominate
presently. While these two problems are separated in quintessence, they 
are degenerate for the cosmological
constant, and one has to introduce the cosmological constant of 
extremely small energy scale at the very beginning of the
universe. 

As a solution of the coincidence problem, the notion of a tracker
field is introduced in \cite{zws}. It is shown that a very wide range
of initial conditions approach a common evolutionary track, so that
the cosmology is insensitive to the initial conditions similar to
inflation. Once one parameter relating to the energy scale of the
vacuum energy is fixed, the present-day equation-of-state 
$w_Q= p_Q/\rho_Q$ is automatically determined: there is a
$\Omega_Q-w_Q$ relation\cite{zws}. 

Direct methods to  verify the idea of quintessence are important. Proposed 
possibilities are the following: the direct reconstruction of the
effective potential from the luminosity distance - redshift relation 
observed for type Ia supernovae\cite{nc}; the detection of 
quintessence from the measurements of a rotation in the plane of
polarization of radiation from  distant radio sources\cite{carroll}.
The direct interaction of the quintessence field to ordinary 
matter, however,  is found to be strongly suppressed so as not to 
violate the equivalence principle and the constancy of the constants of 
nature\cite{carroll}. 

The possibility of an explicit coupling
between the scalar field and the curvature is not excluded
theoretically. It is then natural to consider 
further the coupling of 
the quintessence field to the gravity itself. 
In this paper, we examine the cosmological consequence of the
non-minimal coupling of the quintessence field to the gravity. 
Since such a scalar field gives rise to both the time variation of the 
gravitational constant and a gravity force of long range, 
such a coupling should be constrained by experiments.

\section{Non-minimally Coupled Quintessence}

The action we consider is
\beq
S=\int d^4x\sqrt{-g}\left[{R\over 2\k^2}-{1\over 2}\xi\p^2R-{1\over
  2}g^{ab}\partial_a\p\partial_b\p-V(\p)\right]+S_m,
\label{action}
\eeq
where $\k^2\equiv 8\pi G_{bare}$ is the bare gravitational constant
and $S_m$ denotes the action of matter. 
The effective gravitational ``constant'' is defined by 
$\k^2_{eff}\equiv \k^2 (1-\xi\k^2\p^2)^{-1}$. $\xi$ is the non-minimal 
coupling between the scalar field and the curvature. 
In our conventions, $\xi=1/6$ corresponds to 
the conformal coupling. 

We assume that the universe is described by the flat homogeneous and
isotropic universe model with the scale factor $a$. 
The time coordinate is so normalized that $a=1$ at the present. 
The field equations are then
\beqa
&&H^2\equiv \left({\dot a\over a}\right)^2={\k^2\over 3}
\left[1-\xi\k^2\p^2\right]^{-1}
\left(\rho_B +{1\over 2}\dot\p^2+V(\p)+6\xi H\p\dot\p
\right),\label{eq:1}\\
&&\dot H=-{\k^2\over 2}
\left[1-\xi\k^2\p^2\right]^{-1}
\left[\rho_B+p_B+\dot \p^2+2\xi \left(
H\p\dot\p-\dot\p^2-\p\ddot\p\right)
\right],\label{eq:2}\\
&&\ddot\p+3H\dot\p +6\xi\left(\dot H +2H^2\right)\p
+V'=0,\label{eq:3}\\
&&\dot\rho_B+3H(\rho_B+p_B)=0\label{eq:4},
\eeqa
where $\rho_B, p_B$ denotes the background energy density, pressure,
respectively, and $V'=dV/d\phi$.  

We consider a potential of inverse-power as an example of the
tracker field for $\xi=0$\cite{rp,zws}:
\beq
V(\p)=M^4(\p/M)^{-\alpha}.
\label{potential}
\eeq
For $\xi=0$, there exists
the following scaling solution during the background dominated epoch
\beqa
&&H/H_0= a^{-3(1+w_B)/2} \label{zeroh}\\
&&\p/\p_0= a^{3(1+w_B)/(\alpha +2)} \label{zerop}\\
&&\p_0=\left({2\alpha (\alpha +2)^2 M^{\alpha +4}\over
    9H_0^2(1+w_B)(4+(1-w_B)\alpha)}\right)^{1/(\alpha+2)}.
\eeqa
The  equation-of-state $w_Q$ is
\beq
w_Q={\alpha w_B-2\over \alpha +2}.
\label{eos:zero}
\eeq
Since we consider a potential whose present mass scale is extremely
small ($\siml H_0 \sim 10^{-33}{\rm eV}$), 
the force mediated by the scalar field is of long range, 
and hence the usual solar system limit on $\xi$, likewise the
Brans-Dicke parameter, does apply. The correspondence to the
Brans-Dicke field $\Phi_{BD}$ and the coupling function
$\omega(\Phi_{BD})$ of scalar-tensor theories of gravity\cite{scalar-tensor} 
is given by
\beqa
\Phi_{BD}&=&8\pi(1-\xi\k^2\p^2)/\k^2,\\
\omega(\Phi_{BD})&=&{1-\xi \k^2\p^2\over
  4\xi^2\k^2\p^2}={\k^2\Phi_{BD}\over
  4\xi(8\pi-\k^2\Phi_{BD})}\label{brans}.
\eeqa

\subsection{perturbative analysis}

To consider the effect of $\xi$ qualitatively, we consider the
case of $|\xi| \k^2\p^2 \ll 1$. 
Then during the background dominated epoch, Eq.(\ref{eq:1}) and 
Eq.(\ref{eq:3}) are approximated to
\beqa
&&H^2 = {\k^2\over 3}\rho_B \label{aeq:1}\\
&&\ddot\p+3H\dot\p + \xi\k^2(1-3w_B)\rho_B\p+V'=0,\label{aeq:2}
\eeqa
where we have used Eq.(\ref{eq:2}) to derive Eq.(\ref{aeq:2}). It is
recently established that the scaling solutions
Eq.(\ref{zeroh}-\ref{zerop}) 
with the same power-index persist even if $\xi\neq 0$ and
that the stability of them does not depend on $\xi$\cite{uzan}. 

To the lowest order in $\xi$, the corresponding Brans-Dicke parameter 
is given by  
\beq
\omega_0={1-\xi \k^2\p^2_0\over 4\xi^2\k^2\p^2_0}
\simeq {3\over 4\alpha (\alpha +2)}{1\over \xi^2},
\label{apprx:omega}
\eeq
where we have used the relation that holds for the potential of
inverse-power\cite{zws} to estimate the present-day value of the
scalar field: 
\beq
V''=\alpha(\alpha+1){V\over \p^2}=
{9\over 2}(1-w_Q^2){\alpha+1\over \alpha}H^2.
\eeq
Up to ${\cal O}(\xi)$, the time variation of the gravitational
constant is given by
\beq
{\dot G\over G}\bigg{|}_0={2\xi\k^2\p\dot\p \over 1-
\xi\k^2\p^2}\bigg{|}_0\simeq  2\xi \alpha H_0
\label{gdot:linear} 
\eeq
Hence, for $\xi>0$ the gravitational ``constant'' increases with time, 
while it decreases for $\xi<0$. Eq.(\ref{gdot:linear}) also
shows that $|\dot G/G|$ is larger for larger $\alpha$ since the
potential then becomes steeper and the scalar field rolls down the
potential more rapidly.

\subsection{constraining $\xi$}

We perform the numerical calculation 
to examine in detail the time variation of $G$ and the deviation from
general relativity induced by the non-minimal coupling of 
the quintessence field to the curvature. 
Initial condition is set at $a=10^{-14}$. We vary the fraction
of the energy density of the quintessence field relative to radiation
{}from $10^{-9}$ to $10^{-30}$. We also choose various initial $\p$ and
$\dot\p$. We confirmed the tracking behavior:
convergence to a common evolutionary track\cite{zws,uzan}. 
Below we show typical results for the potential
Eq.(\ref{potential}) with $\alpha=4$. We choose the following
parameters: $\Omega_M\equiv \k_{eff}^2\rho_M/3H^2|_0=0.3$ and 
$H_0=100h{\rm km/sec/Mpc}$ with $h=0.6$. 
 
There exist a lot of experimental limits on the time variation of
$G$\cite{gillies}. Radar ranging data to the Viking landers on Mars
gives $|\dot G/G| = (2\pm 4)\times 10^{-12} {\rm yr}^{-1}$\cite{hellings}. 
Lunar laser ranging experiments yield $|\dot G/G| = (0\pm 11)\times
10^{-12} {\rm yr}^{-1}$\cite{muller} and recently updated as  
$|\dot G/G| = (1\pm 8)\times 10^{-12} {\rm yr}^{-1}$\cite{williams}. 
More recently, a tighter bound is found by analysing the measurements of
the masses of young and old neutron stars in binary pulsars: 
$|\dot G/G| = (0.6\pm 2.0)\times 10^{-12} {\rm yr}^{-1}$\cite{thorsett}, 
although the uncertainties in the age estimation may weaken the
constraint. Considering these experimental results, we will adopt the
limit: $|\dot G/G| = (0\pm 8)\times 10^{-12} {\rm yr}^{-1}$, and the
limit by Thorsett is treated separately.

In Fig. 1, we show the numerical results of $\dot G/G$. 
The shaded region is already excluded 
by the current experimental limits. To examine the model dependencies 
of the results, we also show $\dot G/G$ for the potential of the 
form $M^4\left(\exp(1/\k\p)-1\right)$\cite{zws} by a dotted curve. 
We find that negative $\xi$ is severely constrained, while positive
$\xi$ is loosely constrained and the limit is dependent on the
potential. 

These results are intuitively understood via conformally
transformed picture\cite{fm}.
If we perform the conformal transformation so that the scalar field is 
minimally coupled:
\beq 
g_{ab}=\widetilde{g_{ab}}|1-\k^2\xi\phi^2|^{-1}.
\eeq
Then the action Eq.(\ref{action}) becomes
\beq
S=\int d^4x\sqrt{-\widetilde{g}}\left[{\widetilde{R}\over 2\k^2}-
{1\over 2}F^2(\phi)(\widetilde{\nabla}\phi)^2-\widetilde{V}(\phi)\right]
+S_m,
\eeq
where
\beqa
F^2(\phi)&=& {1-\xi(1-6\xi)\k^2\phi^2\over (1-\xi\k^2\phi^2)^2},\\
\widetilde{V}(\phi)&=&{V(\phi)\over (1-\xi\k^2\phi^2)^2}.
\eeqa
Hence, after redefining the scalar field so that the kinetic term is
canonical:
\beq
\Phi=\int d\phi F(\phi),
\eeq
the action is reduced to that of the scalar field minimally coupled to 
the Einstein gravity. We can follow the dynamics qualitatively by
simply looking at the effective potential $\widetilde{V}(\Phi)$. 
Note that $1/(1-\xi\k^2\phi^2)^2$ is a decreasing function of $\phi$ 
for $\xi<0$, while an increasing function for $\xi>0$. 
For $\xi<0$ the effective potential $\widetilde{V}(\phi)$ decreases
more rapidly than $V(\phi)$ (in particular, $\widetilde{V}(\Phi)$
decreases exponentially for large $\k\p$), and consequently the scalar 
field rolls down the potential more rapidly. On the other hand, 
for $\xi > 0$, $\widetilde{V}(\Phi)$ diverges at $\k^2\p^2 = 1/\xi$, 
so the slope of the effective potential becomes gentler and  
the scalar field rolls down the potential more slowly, and 
hence $|\dot G/G|$ becomes smaller than that for $\xi <0$. 

We may summarize that  the current experimental limits on the time
variation of $G$ constrain the non-minimal coupling as
\beq
- 10^{-2}\siml \xi \siml  10^{-2}\sim 10^{-1},
\eeq
while if the tighter limit by Thorsett is adopted, then we have
\beq
 -10^{-2}\siml \xi \siml 10^{-2}.
\eeq
However, the limit is sensitive to the shape of the potential. 

Most important experimental limits we must consider are the solar
system experiments, such as the Shapiro time delay and 
the deflection of light\cite{will} 
because the non-minimally coupled scalar field can mediate the long range 
gravity force in addition to that mediated by a metric
tensor. The recent experiments set constraint on the
parameterized-post-Newtonian(PPN) parameter $\gamma_{\rm PPN}$ as\cite{data}
\beq
|\gamma_{\rm PPN}-1| < 2\times 10^{-3},
\label{gamma}
\eeq
which constrains  the Brans-Dicke parameter through the relation 
$\gamma_{\rm PPN}=(\omega +1)/(\omega +2)|_0$\cite{will}
\beq
\omega_0 > 500.
\label{omega:500}
\eeq

In Fig. 2, we show the present-day Brans-Dicke parameter defined by
Eq.(\ref{brans}) as a function of $\xi$. We also plot a curve derived
under the assumption of $|\xi|\k^2\p^2\ll 1$, Eq.(\ref{apprx:omega}). 
We find a good agreement. Thus, using Eq.(\ref{apprx:omega}) and
Eq.(\ref{omega:500}),  the non-minimal
coupling $\xi$ is found to be  constrained as 
\beq
|\xi| < 3.9\times 10^{-2}{1\over \sqrt{\alpha(\alpha +2)}} \leq 2.2
  \times 10^{-2}
\eeq
as long as $\alpha \geq 1$. 
The limit is less sensitive to the potential than that derived from
$|\dot G/G|$ 
because $\omega$ does not explicitly depend on $\dot \p$ unlike 
$|\dot G/G|$. We note that the limit is found to be insensitive to 
$\Omega_M$ as long as $\Omega_M \siml 0.7$. 
There is another PPN parameter 
$\beta_{\rm PPN}$ which is written in terms of $\omega$ as 
$\beta_{PPN}-1=(d\omega/d\Phi_{BD})(2\omega +4)^{-1}(2\omega+3)^{-2}|_0$
\cite{will}. The most recent results of the
lunar-laser-ranging\cite{llr}, combined with Eq.(\ref{gamma}), yields
\beq
|\beta_{\rm PPN}-1| < 6\times 10^{-4}.
\eeq
We find that $|\beta_{\rm PPN}-1| \sim {\cal O}(\xi^3)$ and consequently 
the experimental limit on $\beta_{PPN}$ is always satisfied if the
condition Eq.(\ref{gamma}) is satisfied.

\section{summary}

We have explored the possibility of an explicit coupling between the
quintessence field and the curvature. Because the force
mediated by the scalar field is of long range($\simg H^{-1}_0$), 
such a coupling is constrained by the solar system experiments. 
Through both analytical estimate and numerical integration of the 
equations, we have found that the limit is given 
by $|\xi| \siml  10^{-2}$. 
The current limit on the non-minimal coupling, $|\xi| \siml  10^{-2}$, 
is not so strong when compared with other couplings to ordinary
matter. For example, a coupling with the electromagnetic
field is suppressed at the level of $\siml 10^{-6}$; 
the coupling with QCD is at most $\siml 10^{-4}$ \cite{carroll}. 
We have also found that the induced time variation
of $G$ is sensitive to the shape of the potential. The future
improvements in the limit of $\dot G/G$ may further constrain 
negative $\xi$ or might lead to a detection of $\dot G/G <0$ depending 
on the potential of the quintessence field. 

\acknowledgments
The author would like to thank Professor K.Sato for useful comments. 
He also would like to thank the hospitality of Aspen Center for
Physics, where the final part of this work was done. 
This work was supported in part by the JSPS under Grant No.3596.


\begin{figure}
  \begin{center}
  \leavevmode\psfig{figure=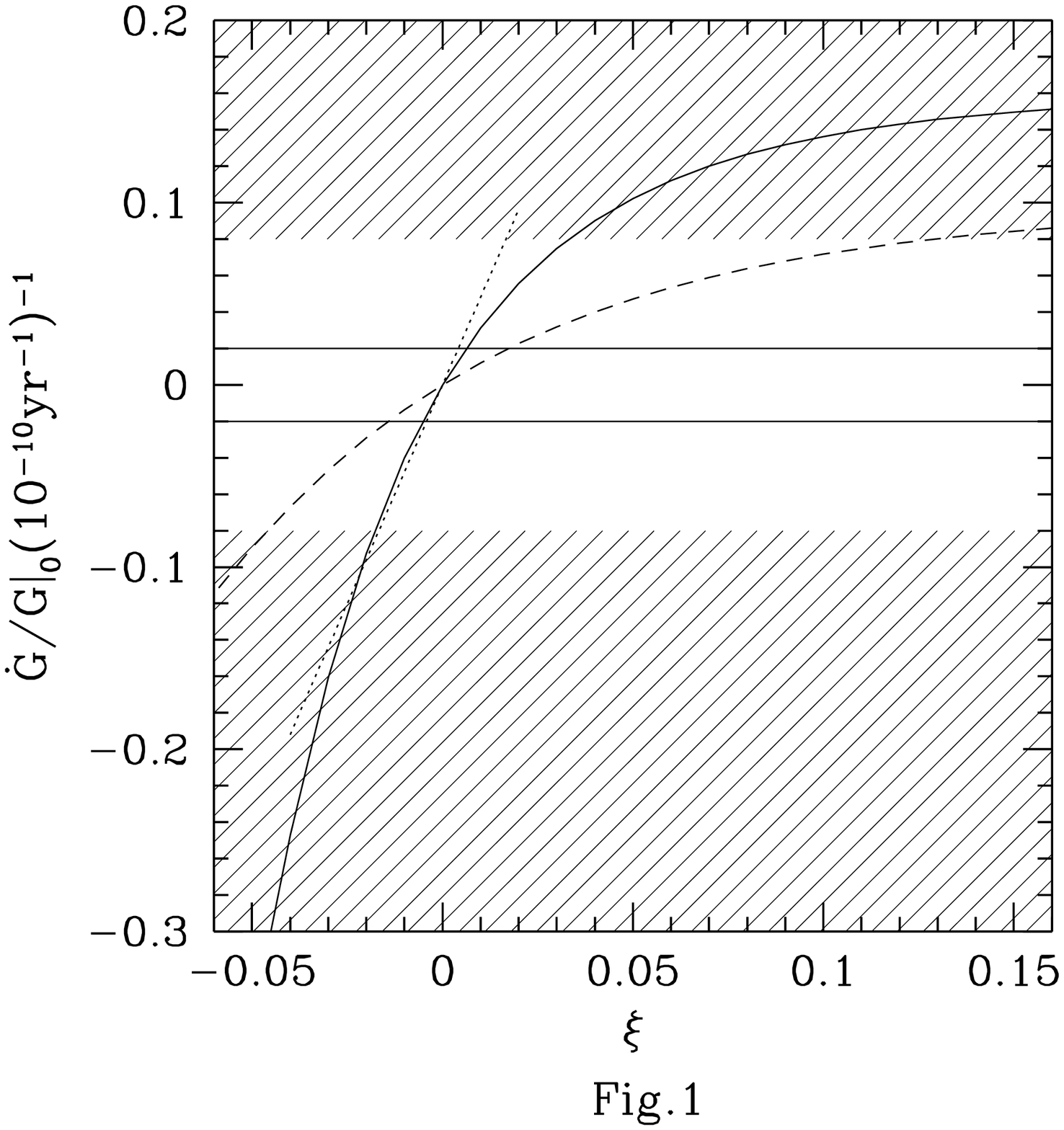,width=15cm}
  \end{center}
  \caption{ The present-day  $\dot G/G$ as a function of $\xi$ for 
the potential of inverse power with $\alpha =4$(solid curve) and for 
the exponential potential(dashed curve).
An approximated relation for $|\xi|\k^2\p^2\ll 1$
(Eq.(\protect\ref{gdot:linear})) is plotted as a dotted line. 
The shaded region is already excluded by the current experimental
limits: $|\dot G/G| = (0\pm 8)\times 10^{-12} {\rm yr}^{-1}$. The
limit by Thorsett is also shown: 
$|\dot G/G| = (0\pm 2)\times 10^{-12} {\rm yr}^{-1}$
\protect\cite{thorsett}.}
  \label{fig:fig1}
\end{figure}

\begin{figure}
  \begin{center}
  \leavevmode\psfig{figure=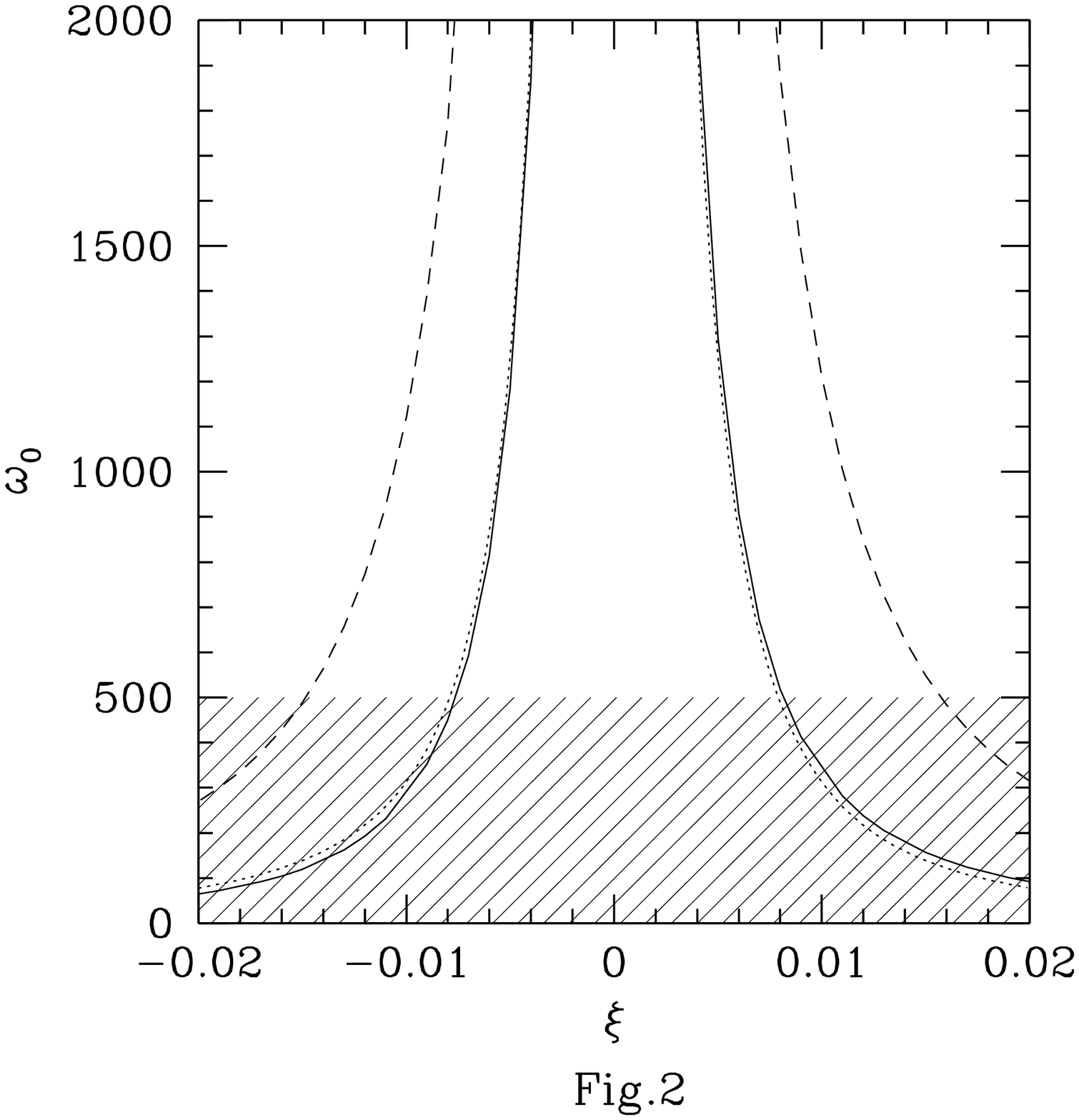,width=15cm}
  \end{center}
  \caption{The present-day Brans-Dicke parameter. The limit by the 
solar system experiments is $\omega_0 > 500$ \protect\cite{will}. 
The solid curve is for the potential of inverse power with $\alpha=4$; 
the dashed curve is for the exponential potential. An approximated
relation for $|\xi|\k^2\p^2\ll 1$ (Eq.(\protect\ref{apprx:omega})) 
is plotted as a dotted curve. 
}
\label{fig:fig2}
\end{figure}


\begin{thebibliography}{123}

\bibitem{os}
J.P. Ostriker and P.J. Steinhardt, Nature, {\bf 377}, 600 (1995); 
L. Wang, R.R. Caldwell, J.P. Ostriker, and P.J. Steinhardt,
astro-ph/9901388. 

\bibitem{rp}
B. Ratra and P.J.E. Peebles, Phys. Rev. D {\bf 37}, 3406 (1988).

\bibitem{fhw}
J.A. Frieman, C.T. Hill, and R. Watkins, Phys. Rev. D {\bf 46}, 1226
(1992).

\bibitem{xmatter}
T. Chiba, N. Sugiyama, and T. Nakamura, MNRAS {\bf 289}, L5 (1997);
MNRAS {\bf 301}, 72 (1998).

\bibitem{tw}
M.S. Turner and M. White, Phys. Rev. D {\bf 56}, 4439 (1997).

\bibitem{cds}
R.R. Caldwell, R. Dave, and P.J. Steinhardt, 
Phys. Rev. Lett. {\bf 80}, 1586 (1998).

\bibitem{bs}
M. Bucher and D. Spergel, astro-ph/9812022.

\bibitem{zws}
I. Zlatev, L. Wang, and P.J. Steinhardt, Phys. Rev. Lett. {\bf 82}, 896
(1999); P.J. Steinhardt, L. Wang, and I. Zlatev, Phys. Rev. D {\bf
  59}, 123504 (1999).

\bibitem{nc}
T. Nakamura and T. Chiba, MNRAS {\bf 306}, 696 (1999); 
T. Chiba and T. Nakamura, Prog. Theor. Phys., {\bf 100},
  1077 (1998); A. Starobinsky, JETP Lett. {\bf 68}, 757 (1998); 
D. Huterer and M.S. Turner, astro-ph/9808133.

\bibitem{carroll}
S.M. Carroll, Phys. Rev. Lett. {\bf 81}, 3067 (1998).

\bibitem{scalar-tensor}
C. Brans and R.H. Dicke, { Phys. Rev.} {\bf 124}, 925 (1961); 
P.G. Bergmann, { Int. J. Theor. Phys.} {\bf 1}, 25 (1968);
K. Nordtvedt, { Astrophys. J.} {\bf 161}, 1059 (1970);
R.V. Wagoner, { Phys. Rev.} {\bf D1}, 3209 (1970).

\bibitem{uzan}
J.-P. Uzan, Phys. Rev. D {\bf 59}, 123510 (1999).

\bibitem{gillies}
G.T. Gillies, Rep. Prog. Phys. {\bf 60}, 151 (1997).

\bibitem{hellings}
R.W. Hellings et al., Phys. Rev. Lett. {\bf 51}, 1609 (1983).

\bibitem{muller}
J. M\"uller, M. Schneider, M. Soffel, and H. Ruder, 
Astrophys.J. {\bf 382}, L101 (1991).

\bibitem{williams}
J.G. Williams, X.X. Newhall, and J.O. Dickey, Phys. Rev. D {\bf 53}, 6730
(1996).

\bibitem{thorsett}
S.E. Thorsett, Phys. Rev. Lett. {\bf 77}, 1432 (1996).

\bibitem{fm}
T. Futamase and K. Maeda, Phys. Rev. D {\bf 39}, 399 (1989).

\bibitem{will}
C.M. Will, {\it Theory and experiment in gravitational physics},
(Rev.Ed., Cambridge University Press, Cambridge, 1993).

\bibitem{data}R. D. Reasenberg et al.,
{ Astrophys.J.} {\bf 234}, L219 (1979);
R.S. Robertson, W.E. Carter, and
W.H. Dillinger, { Nature} {\bf 349}, 768 (1991).

\bibitem{llr}
J.G. Williams, X.X. Newhall, and J.O. Dickey, 
{Phys. Rev. D} {\bf 53}, 6730 (1996).


\end{thebibliography}
\end{document}